\documentclass[a4paper,12pt]{article}

\usepackage[a4paper]{geometry}
\usepackage[dvipsnames]{xcolor}
\usepackage[utf8]{inputenc} 
\usepackage[hidelinks]{hyperref}
\usepackage{float}
\usepackage{amssymb,amsmath,amsthm}
\usepackage{amsfonts}

\pdfoutput=1
\usepackage{graphicx}
\usepackage{graphicx, float, tabularx, booktabs}
\usepackage{color}
\usepackage{enumerate}
\usepackage[american]{babel}
\usepackage{stackrel}

\usepackage[numbers, compress]{natbib}
\usepackage{natbib}
\newcommand{\bq}{\begin{eqnarray}}
\newcommand{\fq}{\end{eqnarray}}
\newcommand{\be}{\begin{equation}}
\newcommand{\ee}{\end{equation}}
\newcommand{\bea}{\begin{eqnarray}}
\newcommand{\eea}{\end{eqnarray}}

\usepackage{fancyhdr}
\usepackage{amsfonts}
\usepackage{epsfig}

\usepackage{graphicx, tabularx}
\newcommand{\beq}{\begin{equation}}
\newcommand{\eeq}{\end{equation}}

\vspace{-0.1cm}
\renewcommand{\thefootnote}{\fnsymbol{footnote}}

\begin{document}

\begin{center}
{\Large \textbf{On the Lichnerowicz operator in traversable wormhole
spacetimes}}
\end{center}

\vspace{-0.1cm}

\begin{center}
Remo Garattini, $^{a,b}$\footnote{%
remo.garattini@unibg.it} and Piero Nicolini $^{a,c,d}$\footnote{%
nicolini@fias.uni-frankfurt.de}

\vspace{.6truecm}

\emph{$^a$Dipartimento di Ingegneria e Scienze Applicate,\\[0pt]
Università degli Studi di Bergamo,\\[0pt]
Viale Marconi, 5, 24044 Dalmine (Bergamo), Italy }\\[0pt]

\emph{$^b$INFN, Sezione di Milano, Milan, Italy}\\[0pt]

\emph{$^c$Frankfurt Institute for Advandced Studies (FIAS)\\[0pt]
Ruth-Moufang-Str. 1, 60438 Frankfurt am Main, Germany}\\[0pt]

\emph{$^d$Institut f\"ur Theoretische Physik,\\[0pt]
Johann Wolfgang Goethe-Universit\"at Frankfurt am Main\\[0pt]
Max-von-Laue-Str. 1 60438 Frankfurt am Main, Germany}\\[0pt]
\end{center}

\begin{abstract}
%\nin TO BE UPDATED
\noindent{\small The evaluation of Casimir energies in curved background spacetimes is an essential ingredient to
study the stability of  traversable wormholes. 
In practice one has to calculate the contribution of the transverse-traceless component of the metric perturbation on a curved spacetime background. This implies the study of an eigenvalue equation
involving a modified form of the Lichnerowicz operator. For arbitrary background spacetimes, however, such
an operator does not display transverse-traceless properties, a fact that
impedes the determination of the eigenvalues. Against this background, we show
that the problem can be
circumvented. Casimir energies can be calculated by gauging the original form of the modified Lichnerowicz operator into a transverse-traceless one. } \noindent

\end{abstract}

%\maketitle

\renewcommand{\thefootnote}{\arabic{footnote}} \setcounter{footnote}{0}
%\thispagestyle{empty}
%\clearpage
%\tableofcontents

\section{Introduction}

Traversable wormholes are spacetime geometries emerging as solutions of
Einstein equations. Conventionally they are thought as fascinating
configurations that might have formed in remote regions of the Universe.
Nowadays such a perspective has changed: The understanding of the physics
governing wormholes is instrumental for the concrete realization of laboratory
devices for interstellar travels \cite{White03}.

The current efforts in such a research field  are focused on the
conditions of traversability of wormholes. In practice, the wormhole throat
can be stable only if there is a source of energy able to balance the
gravitational pull. Contrary to the case of ordinary stars,  such a balance  implies the source to be of
\textit{exotic} type. The latter is a term used to indicate a non standard
matter that violates the null energy conditions (NEC), namely the positiveness
of the energy momentum tensor, $T_{\mu\nu}k^{\mu}k^{\nu}\geq0$ for any null
vector $k^{\mu}$. Wormholes necessitate such a violation of NEC since, according to the Raychaudhuri equation \cite{Ray55}, the expansion of timelike congruence becomes negative at the throat while remains positive  elsewhere. In other words, a set of world lines undergoes a contraction at the throat, indicating an inevitable collapse into a singularity unless exotic matter  locally  counteracts it. Although the pressure of ordinary matter can, in general, counteract a collapse, it would not be enough high to reestablish the positiveness of the congruence expansion on the other side of the throat \cite{MoT88}.  As far as we know, the Casimir energy represents the only
artificial source of exotic matter realizable in a laboratory \cite{White03,Garattini:1999mh,Garattini:2000kd,Garattini19,Garattini20}. Alternatively, a local violation of energy conditions can   occur due to the quantum fluctuation of the graviton. This fact propelled 
the idea of \textit{self-sustained traversable wormholes}, that have been introduced
in \cite{Garattini05,Garattini07c,Garattini:2010dn}. To study such wormholes one considers the semiclassical Einstein equations, 
\begin{equation}
G_{\mu\nu}=\kappa \left<T_{\mu\nu}\right>^\mathrm{ren}, \quad \left<T_{\mu\nu}\right>^\mathrm{ren}=-\frac{1}{\kappa}\left<\Delta G_{\mu\nu}\left(\bar{g}_{\mu\nu}, h_{\mu\nu}\right)\right>^\mathrm{ren} 
\label{eq:semclEE}
\end{equation} 
where the source term is the expectation value of the renormalized quantum stress tensor of the metric perturbation $h_{\mu\nu}$, with $g_{\mu\nu}=\bar{g}_{\mu\nu}+h_{\mu\nu}$. Eq. \eqref{eq:semclEE} simplifies by focusing on the energy components, namely by  a projection on a constant time spacelike hypersurface $\Sigma$. In such a way, one obtains Hamiltonian and energy densities, that, after integration, give the equation for the stability of the wormhole:%
\begin{equation}
H_{\Sigma}^{(0)}=-E^{\bot}.\label{SS}%
\end{equation}
Here $E^{\bot}$ is the total regularized graviton one loop energy coming from the quantized stress tensor and
$H_{\Sigma}^{(0)}$ is the classical term, coming from the Einstein tensor. Only the transverse traceless (TT)
component of the graviton contribute to the energy $E^{\bot}$ and for this
reason we introduced  the superscript $^{\bot}$. 
For a
spherically symmetric line element of the form%
\begin{equation}
ds^{2}=-\exp\left(  -2\Phi\left(  r\right)  \right)  dt^{2}+\frac{dr^{2}%
}{1-\frac{b\left(  r\right)  }{r}}+r^{2}d\Omega^{2},\label{metric}%
\end{equation}
where $\Phi\left(  r\right)  $ is the redshift function, $b\left(  r\right)  $
is the shape function and $d\Omega^{2}=d\theta^{2}+\sin^{2}\theta d\phi^{2}$
is the line element of the unit sphere, the classical term reduces
to%
\begin{align}
H_{\Sigma}^{(0)} &  =\int_{\Sigma}\,d^{3}x\left[  \,\left(  16\pi G\right)
G_{ijkm}\pi^{ij}\pi^{km}-\frac{\sqrt{g}}{16\pi G}\!{}\!\,\ ^3R\right]
\nonumber\\
&  =-\frac{1}{16\pi G}\int_{\Sigma}\,d^{3}x\,\sqrt{g}\,\  ^3R\,=-\frac{1}{2G}%
\int_{r_{0}}^{\infty}\,\frac{dr\,r^{2}}{\sqrt{1-b(r)/r}}\,\frac{b^{\prime}%
(r)}{r^{2}}\,.
\end{align}
Here we have expressed the three dimensional scalar
curvature  $^3R$  in terms of $b(r)$. The symbol
$G_{ijkm}$ denotes the super-metric and $\pi^{ij}$ the super-momentum. Due to static conditions the kinetic term  $G_{ijkm}\pi^{ij}\pi^{km}$ disappears. For a full derivation see \cite{GaL17}. 
At the one-loop
approximation level, the energy $E^{\bot}$  is identified as a Casimir like energy
 against the fixed background. Its evaluation requires functional integration methods including the solution of
%$h_{ij}$
%\bq
%g_{ij}=\bar{g}_{ij}+h_{ij},
%\fq
%Such a graviton energy can be evaluated by solving 
an appropriate eigenvalue
equation in terms of  a modified Lichnerowicz operator (see \cite{GaL17} for the full derivation).

 The conventional Lichnerowicz operator can be defined through its
action on a tensor $h_{ij}$ as 
\begin{align}
&  \left(  \bigtriangleup_{L}h\right)  _{ij}=\bigtriangleup h_{ij}%
-2R_{ikjm}h^{km}+R_{ik}h_{j}^{k}+R_{jk}h_{i}^{k}\nonumber\\
&  \bigtriangleup=-\nabla^{a}\nabla_{a},\label{DeltaL}%
\end{align}
where Latin indexes run from $1$ to $3$ and $\nabla^a$ is the covariant derivative with respect to the $3$-metric,  $\bar{g}_{ij}$. For ease of notation we employ, in the following equations, the symbol $g_{ij}$ for the background
without superscript $\bar{{}}$. 
As said above, we have to 
%In order to describe the graviton contribution, we assume $h_{ij}$ to be the
%perturbation of the background metric and we 
consider the TT 
component of the field $h_{ij}$ describing a spin 2 particle, namely
\begin{equation}
g^{ij}h_{ij}^{\bot}=0,\qquad\nabla^{i}h_{ij}^{\bot}=0.
\end{equation}
 Thus, the problem turns into the
determination of the eigenvalues of the following modified Lichnerowicz
operator,
\begin{equation}
\left(  \tilde{\bigtriangleup}_{L\!}\!{}h^{\bot}\right)  _{ij}=\left(
\bigtriangleup_{L\!}\!{}h^{\bot}\right)  _{ij}-4R{}_{i}^{k}\!{}\ h_{kj}^{\bot
}+\text{ }^{3}R{}\!{}\ h_{ij}^{\bot}.\label{M Lichn}%
\end{equation}
namely
\begin{equation}
\left(  \tilde{\bigtriangleup}_{L\!}\!{}h^{\bot}\right)  _{ij}=\lambda
\!{}h_{ij}^{\bot},\label{Eq}%
\end{equation}
describing the energy spectrum of $h_{ij}^{\bot}$  resulting from \eqref{eq:semclEE}.
To have a well-posed equation in \eqref{Eq}, one has, however, to make sure that the
above operator does not change the TT properties of $h_{ij}^{\bot}$. This is, in general, a major issue because the l.h.s. is not a TT tensor for some kind of backgrounds \cite{Delay07}.
%Indeed, the Lichnerowicz operator and its modified version are not TT tensors
%\cite{Delay07}. 

The present paper aims to circumvent such difficulties  and pave the way of a consistent study of wormhole
stability within this formalism.

\section{The Lichnerowicz operator for TT tensors}

\label{appT}
%Our starting point is the modified Lichnerowicz operator in \eqref{M Lichn}. We aim to show upon what conditions such an operator can still have TT properties.

It is straightforward to see that the standard Lichnerowicz operator, $\left(
\bigtriangleup_{L}h^{\bot}\right)  _{ij}$, is traceless. This is, however, not
enough to conclude that $\left(  \tilde{\bigtriangleup}_{L\!}\!{}h^{\bot}\right)  _{ij}$ is traceless too.
Indeed, one has
\begin{equation}
\mathrm{Tr}\left[  \left(  \bigtriangleup_{L}h^{\bot}\right)  _{ij}%
-4R_{ik}h_{j}^{{\bot}\ k}+\text{ }^{3}Rh_{ij}^{\bot}\right]  =-4R_{k}^{m}%
h_{m}^{{\bot}\ k}.\label{Tr}%
\end{equation}
As a first step, we aim to write such an operator in terms of a trace free part
and a term determining the trace. For the line
element $\left(  \ref{metric}\right)  $, the mixed Ricci tensor $R_{j\text{ }%
}^{i}$ is:
\begin{equation}
R_{j}^{i}=\left\{  \frac{b^{\prime}\left(  r\right)  }{r^{2}}-\frac{b\left(
r\right)  }{r^{3}},\frac{b^{\prime}\left(  r\right)  }{2r^{2}}+\frac{b\left(
r\right)  }{2r^{3}},\frac{b^{\prime}\left(  r\right)  }{2r^{2}}+\frac{b\left(
r\right)  }{2r^{3}}\right\}  .\label{eq:riccimixed}%
\end{equation}
Therefore, we can write
\[
R_{k}^{i}h_{j}^{k\ \bot}=\underset{\mathrm{traceless}}{\underbrace{f\left(
r\right)  h_{i}^{\bot\ j}}}+\left(  R_{1}^{1}-f\left(  r\right)  \right)
\delta_{j}^{1}\delta_{1}^{i}h_{i}^{\bot\ j}%
\]
where we have used the properties $R_{2}^{2}=R_{3}^{3}\equiv f\left(
r\right)  $ and $h_{1}^{\bot\ 1}+h_{2}^{\bot\ 2}+h_{3}^{\bot\ 3}=0$. As a
result, one finds
\begin{equation}
\left(  \tilde{\bigtriangleup}_{L\!}\!{}\ h^{\bot}\right)  _{i}^{j}=\left(
\tilde{\bigtriangleup}_{L\!}\!{}\ h^{\bot}\right)  _{i}^{j\ \mathrm{T}}%
+
\left\{
\frac{1}{2}\left[  \frac{3b\left(  r\right)  }{r^{3}}-\frac{b^{\prime}\left(
r\right)  }{r^{2}}
\right]
  \delta_{j}^{1}\delta_{1}^{i}
  \right\}
h_{i}^{\bot
\ j}\label{eq:trace}%
\end{equation}
with the trace free part defined as
\begin{equation}
\left(  \tilde{\bigtriangleup}_{L\!}\!{}h^{\bot}\right)  _{i}^{j\ \mathrm{T}%
}\equiv\left(  \bigtriangleup_{L}h^{\bot}\right)  _{i}^{j}-4f\left(  r\right)
h_{i}^{\bot\ j}+\text{ }^{3}Rh_{i}^{\bot\ j}.
\label{eq:tracefreepart}
\end{equation}
Now we compute the divergence of the above trace free part,
namely \cite{Delay07}
\begin{align}
& \nabla^{m}\left(  \tilde{\bigtriangleup}_{L\!}\!{}\ h^{\bot}\right)
_{m}^{j\ \mathrm{T}}=\nabla^{m}\left(  \bigtriangleup_{L}h^{\bot}\right)
_{m}^{j}-\nabla^{m}\left(  4f\left(  r\right)  h_{m}^{{\bot}\,j}-\text{ }%
^{3}Rh_{m}^{{\bot}\,j}\right) \nonumber\\
& =\bigtriangleup\left(  \nabla^{m}h_{m}^{{\bot}\,j}\right)  +R_{k}^{j}\left(
\nabla^{m}h_{m}^{{\bot}\,k}\right)  +\left(  \nabla^{j}R_{m}^{k}\right)
h_{k}^{m\,{\bot}}-\nabla^{m}\left(  4f\left(  r\right)  h_{m}^{{\bot}%
\,j}-\text{ }^{3}Rh_{m}^{{\bot}\,j}\right) \nonumber\\
& =\left(  \nabla^{j}R_{m}^{k}\right)  h_{k}^{\bot m}-4\left(  \nabla
^{m}f\left(  r\right)  \right)  h_{m}^{\bot\,j}+\left(  \nabla^{m}\text{ }%
^{3}R\right)  h_{m}^{\bot\,j}.\label{eq:divtracefree}%
\end{align}
where we have used the transverse property $\nabla^{m}h_{m}^{\bot\,j}=0$ and
$$\nabla^{m}\left(\bigtriangleup_{L\!}\!{}\ h^{\bot}\right)_{m}^{j\ \mathrm{T}}=\bigtriangleup\left(  \nabla^{m}h_{m}^{{\bot}\,j}\right)  +R_{k}^{j}\left(
\nabla^{m}h_{m}^{{\bot}\,k}\right)  +\left(  \nabla^{j}R_{m}^{k}\right)
h_{k}^{m\,{\bot}}.$$ In
summary one finds:%
\begin{equation}
\nabla^{m}\left(  \tilde{\bigtriangleup}_{L\!}\!{}h^{\bot}\right)
_{m}^{j\ \mathrm{T}}=\left\{-\frac{1}{2}\ g^{11}\partial_{1}\left[  \frac{7b\left(
r\right)  }{r^{3}}-\frac{b^{\prime}\left(  r\right)  }{r^{2}}\right]
\delta_{j}^{1}\delta_{1}^{m}\right\} h_{m}^{j\,\bot}.\label{eq:tracefreeLop}%
\end{equation}
The r.h.s. of the above equation vanishes provided
\[
\frac{7b\left(  r\right)  }{r^{3}}-\frac{b^{\prime}\left(  r\right)  }{r^{2}%
}=\mathrm{constant}.
\]
Thus, one obtains
\begin{equation}
b(r)=Ar^{3}
\label{eq:br}
\end{equation}
where $A$ is a constant coefficient. Accordingly, one has $A>0$ for de Sitter
space, $A<0$ for Anti-de Sitter space and $A=0$ for Minkowski space. In such
cases, one finds that the trace in \eqref{eq:trace} vanishes. As a result, one
can conclude, from the condition \eqref{eq:br}, that the 
operator, $\left(  \tilde{\bigtriangleup}_{L\!}\!{}h^{\bot}\right)  _{ij}$, is a TT tensor in case of constant or vanishing curvature.

In the presence of a gravitational source, the curvature is, in general, not
constant. The operator $\left(  \tilde{\bigtriangleup}_{L\!}\!{}h^{\bot}\right)  _{ij}$ can, however, satisfy the
transverse condition up to negligible terms if the curvature variation is
small. The reference scale in such a case is the Planck mass cubed,
$M_{\mathrm{P}}^{3}$. The vanishing of the trace in \eqref{eq:trace} requires
the curvature itself to be small with respect to $M_{\mathrm{P}}^{2}$, namely
\begin{align}
\label{eq:curvcond}\frac{b}{r}\ll\frac{r^{2}}{L_{\mathrm{P}}^{2}},\quad
b^{\prime}\ll\frac{r^{2}}{L_{\mathrm{P}}^{2}},
\end{align}
where $L_{\mathrm{P}}=1/M_{\mathrm{P}}$. As a result, the trace freedom is a
stronger condition with respect to the small variation of the curvature. Both
conditions are easily met in the large distance limit. It is sufficient to
assume $r\gg L_{\mathrm{P}}$ for having the TT condition fulfilled, provided
$b^{\prime}$ is bounded.

At short distance, namely at the wormhole throat, the r.h.s. of
\eqref{eq:tracefreeLop} vanishes because of the presence of the metric
coefficient $g^{11}$. The conditions \eqref{eq:curvcond} are fulfilled for a
wormhole throat $r_\mathrm{t}$ such that $r_\mathrm{t}\gg L_{\mathrm{P}}$,
being $b^\prime(r_\mathrm{t})<1$ according to the flaring out condition.

Up to now, the presented analysis has been focused on differential
conditions. %, namely conditions that have to be fulfilled at specific points of the manifold.
 To calculate the sought eigenvalues, one has, however,
to consider matrix elements, expressed in terms of the following integral%
\begin{equation}
\int_{\Sigma}d^{3}x\sqrt{g}\ h_{j}^{\bot\,i}\left(  \tilde{\bigtriangleup}%
_{L}h^{\bot}\right)  _{i}^{j},\label{eq:matrixel}%
\end{equation}
resulting from the one loop Hamiltonian at the r.h.s of \eqref{eq:semclEE}, 
\begin{align}
&  H_{\Sigma}^{\bot}=\frac{1}{4V}%
\int_{\Sigma}d^{3}x\sqrt{\bar{g}}\ G^{ijkm}\left[  \left(  2\kappa\right)
K^{-1\bot}\left(  x,x\right)  _{ijkm}%\right.  \nonumber\\ &  \left. 
 +\frac{1}{\left(  2\kappa\right)  }\!{}\left(  \tilde
{\bigtriangleup}_{L\!}\right)  _{j}^{a}K^{\bot}\left(  x,x\right)
_{iakm}\right], \label{p22}%
\end{align}
where  
 \begin{equation}
K^{\bot}\left(  \overrightarrow{x},\overrightarrow{y}\right)  _{iakm}%
=\sum_{\tau}\frac{h_{ia}^{\left(  \tau\right)  \bot}\left(  \overrightarrow
{x}\right)  h_{km}^{\left(  \tau\right)  \bot}\left(  \overrightarrow
{y}\right)  }{2\ell\left(  \tau\right)  },\label{proptt}%
\end{equation}
is the graviton propagator, $\ell(\tau)$ a set of variational parameters to be determined by minimizing \eqref{p22} and $\tau$ denotes a complete set of indexes (see \cite{GaL17}).

One can try to circumvent the problem of the TT nature of the operator $\left(  \tilde{\bigtriangleup}_{L\!}\!{}h^{\bot}\right)  _{ij}$, since integral relations generally demand
softer conditions than differential relations. To this purpose we observe that
%the above integral can be written as,
\begin{align}
\int_{\Sigma}d^{3}x\sqrt{g}\ h_{j}^{\bot\,i}\left(  \tilde{\bigtriangleup}%
_{L}h^{\bot}\right)  _{i}^{j}  & =\int_{\Sigma}d^{3}x\sqrt{g}\ h_{j}^{\bot
\,i}\left[  \left(  \tilde{\bigtriangleup}_{L}h^{\bot}\right)  _{i}^{j}%
+\frac{4}{3}\ \delta_{i}^{\ j}\left(  R_{m}^{\ k}\ h_{k}^{\bot\ m}\right)
\right]  =\nonumber\\
& =\int_{\Sigma}d^{3}x\sqrt{g}\ h_{j}^{\bot\,i}\left(  \tilde{\bigtriangleup
}_{L\!}\!{}\ h^{\bot}\right)  _{i}^{j\ \mathrm{T}}.\label{Ip}%
\end{align}
The integral of the term $R_{m}^{\ k}\ h_{k}^{\bot\ m}$ at the r.h.s.
identically vanishes, being $\mathrm{Tr}\ h_{k}^{\bot\ m}=h_{k}^{\bot\ k}=0 $.
Due to \eqref{Tr} and the generic relation
\begin{equation}
\left(  T_{ij}\right)  ^{\mathrm{T}}=T_{ij}-\frac{1}{3}g_{ij}\left[
\mathrm{Tr}\ T_{km}\right]  .
\end{equation}
we obtained the trace free part. The 
operator $\left(  \tilde{\bigtriangleup}_{L\!}\!{}h^{\bot}\right)  _{ij}$ is not traceless, but its integral is equivalent to that of its trace
free part. This property is instrumental to prove that, at level of integral relations,  the transverse property is 
satisfied too.  After gauging the trace away by means of \eqref{Ip}, the transverse property  can be analyzed by considering just the integral of  the trace free part \eqref{eq:tracefreepart}. The latter can be written as
\begin{equation}
\left(  \tilde
{\bigtriangleup}_{L}h^{\bot}\right)  _{ij}^{\mathrm{T}}\equiv \left(  \bigtriangleup
_{L}h^{\bot}\right)  _{ij}-4\left(  R_{ik}h_{j}^{{\bot}\ k}-\frac{1}%
{3}\ g_{ij}\left(  R_{m}^{\ k}h_{k}^{\bot\ m}\right)  \right)  +\text{ }%
^{3}Rh_{ij}^{\bot}.
\end{equation}
We can further gauge the integral of  $\left(  \tilde{\bigtriangleup}%
_{L}h^{\bot}\right)  _{i}^{j\ \mathrm{T}}$ by adding a vanishing contribution whose integrand is  traceless in order not to  alter the trace free property, namely
\begin{equation}
\label{eq:inttransversegauge}
\int_{\Sigma}d^{3}x\sqrt{g}\ h_{j}^{\bot\,i}\left(  \tilde{\bigtriangleup}%
_{L}h^{\bot}\right)  _{i}^{j\ \mathrm{T}}=\int_{\Sigma}d^{3}x\sqrt{g}%
\ h_{j}^{\bot\,i}\left[  \left(  \tilde{\bigtriangleup}_{L}h^{\bot}\right)
_{i}^{j\ \mathrm{T}}+\left(  LM\right)  _{i}^{j}\right]  ,
\end{equation}
where
\begin{equation}
\left(  LM\right)  _{ij}=\nabla_{i}M_{j}+\nabla_{j}M_{i}-\frac{2}{3}%
g_{ij}\left(  \nabla_k M^k\right).
\label{eq:LM}
\end{equation}
One can  check that the integral of $\left(  LM\right)  _{ij}$ is vanishing  by
integrating by parts the first two terms $\nabla_{i}M_{j}$   and using the
transverse condition $\nabla_{i}h_{j}^{\bot\,i}=0$. The third term vanishes
because $h_{j}^{\bot\,i}$ is trace free.

At this point, one can suitably select  $\left(  LM\right)  _{ij}$ 
to get the tranverse condition, namely%
\begin{eqnarray}
&& 0   =\nabla^{i}\left\{  \left[  \left(  \bigtriangleup_{L}h^{\bot
}\right)  _{ij}-4\left(  R_{im}h_{j}^{{\bot}\ m}-\frac{1}{3}\ g_{ij}\left(
R_{i}^{\ m}\ h_{m}^{\bot\ i}\right)  \right)  +\text{ }^{3}Rh_{ij}^{\bot
}\right]+\left(  L M\right)  _{ij}\right\}  =\nonumber\\
&& =\left(  \nabla_{j}R_{i}^{m}\right)  h_{m}^{\bot i} - \nabla^{i}\left[4 R_{im}h_{j}^{{\bot}\ m}-\frac{4}{3}\ g_{ij}\left(
R_{i}^{\ m}\ h_{m}^{\bot\ i}\right)-\text{ }^{3}Rh_{ij}^{\bot
}-\left(  L M\right)  _{ij}\right] =\nonumber
 \\
% && =\nabla^{i}\left[  \left(  \left(  \bigtriangleup_{L}h^{\bot}\right)_{ij}-4\left(  Rh^{{\bot}}\right)  _{ij}+\text{ }^{3}h_{ij}^{\bot}\right)^{\mathrm{T}}+\left(  L\cdot M\right)  _{ij}\right]  \qquad\qquad\left(Rh^{{\bot}}\right)  _{ij}\nonumber\\
% &&=R_{il}h_{j}^{{\bot}\ l}-\frac{1}{3}\ g_{ij}\left( R_{i}^{\ l}\ h_{l}^{\bot\ i}\right)\nonumber \\
%  && =\left(  \nabla^{j}R_{i}^{l}\right)  h_{l}^{\bot i}-4\nabla^{i}\left( Rh^{{\bot}}\right)  _{ij}+\nabla^{i}\left(  \text{ }^{3}Rh_{ij}^{\bot}\right) +\nabla^{i}\left(  L\cdot M\right)  _{ij}. \nonumber\\
% && \left(  \nabla^{j}R_{i}^{l}\right)  h_{l}^{\bot i}-\nabla^{i}\left( 4Rh^{{\bot}}-L\cdot M-\ ^{3}Rh^{\bot}\right)  _{ij}.
&&=\left(  \nabla_{j}R_{i}^{m}\right)  h_{m}^{\bot i}-4\left(  \nabla
^{i}f\left(  r\right)  \right)  h_{ij}^{\bot}+\left(  \nabla^{i}\text{ }%
^{3}R\right)  h_{ij}^{\bot}+\nabla^{i}\left(  LM\right)  _{ij}.\label{Mel}%
\end{eqnarray}
provided \eqref{Mel} has solutions.

Alternatively one can consider, in place of $(LM)_{ij}$ in \eqref{eq:inttransversegauge}, an antisymmetric term of the kind
\begin{equation}
\left( L N\right) _{ij}=\nabla_{i}N_{j}-\nabla_{j}N_{i},
\end{equation}
that is trace free and has a vanishing integral. Its covariant derivative is
formally equivalent to the four current of the electromagnetic field tensor,
namely
\begin{equation}
\nabla^{i}\left( L N\right) _{i}^{j}=S^{j}.
\end{equation}
In such a case, one can select  $S^{j}$ provided the equivalent of \eqref{Mel} for $(LN)_{ij}$ has solutions.

In conclusion, % the sought eigenvalues can be calculated by ``gauging'' the modified Lichnerowicz operator. 
even if the operator is not a TT tensor
for arbitrary backgrounds, its integral \eqref{eq:matrixel} is
equivalent to the integral of an operator \eqref{eq:inttransversegauge} that display TT properties.

\section{Final remarks}

In this paper we have presented a solution to an open issue in the literature,
namely the calculation of graviton energies at the one-loop
approximation associated to a Lichnerowicz operator. %Interestingly  such a problem is present even in Quantum Cosmology when the Wheeler-De Witt equation is considered to estimate the cosmological constant. 
In case of
spherically symmetric spacetimes, such energies
come from the TT component of the perturbation, namely
\begin{equation}
E^{\bot}=\frac{1}{3}\sum_\tau\left[\sqrt{\lambda_1^2(\tau)}+\sqrt{\lambda_2^2(\tau)}\right]
\end{equation}
where the eingenvalues  correspond to the two graviton polarizations.
 Unfortunately, the operator $\left(  \tilde{\bigtriangleup}_{L\!}\!{}h^{\bot}\right)  _{ij}$ is not, in general, a TT tensor,
% \textbf{and the related eigenvalues cannot be determined. This} 
a fact that deprives the
formalism of its predictive power, apart from the case of specific spacetimes
 where the eigenvalue equations can be solved.

Against this background, %In such a framework, our results contribute to fill a gap in the literature.
we have shown that $E^{\bot}$ can be calculated in terms of another
operator that exhibits TT properties. Such an operator is obtained by a
suitable ``gauge'' of the original  operator $\left(  \tilde{\bigtriangleup}_{L\!}\!{}h^{\bot}\right)  _{ij}$ in the integral relations \eqref{eq:matrixel} and \eqref{p22}.

The proposed results can pave the way to further studies based on the
Lichnerowicz operator to scrutinize the conditions of stability of traversable
wormholes spacetimes.

\subsection*{Acknowledgements}

This work has been supported by the project ``Traversable Wormholes: A Road To Interstellar Exploration,'' an Interstellar Initiatives Grant award funded by the Limitless Space Institute.
% This work has been supported by the project ``Traversable Wormholes: A Road To Interstellar Exploration'' of Texas A \& M Engineering Experiment Station (TEES), The Texas A \& M University System. 
The work of P.N. has partially
been supported by GNFM, the Italian National Group for Mathematical Physics.
%The  work of  A.G.T.  has  been  supported  by  the  GRADE  Completion  Scholarships,   which  are  funded  by  the  STIBET  program  of  the  German  Academic  Exchange  Service  (DAAD)  and  the  Stiftung  zur  Förderung  der  internationalen wissenschaftlichen Beziehungen der Johann Wolfgang Goethe-Universität.
%\iffalse


\end{document}